\documentclass{nature}
\usepackage{amssymb,amsmath}
\usepackage{graphicx}
\usepackage[breaklinks=true,bookmarksopen=true,bookmarksopenlevel=3,bookmarksnumbered=true,colorlinks=true,urlcolor= magenta,citecolor=blue,linkcolor=blue]{hyperref}

\begin{document}

\title{Demonstration of relativistic electron beam focusing by  a laser-plasma lens}

\author{C. Thaury$^1$, E. Guillaume$^1$, A. D\"opp$^{1,2}$, R. Lehe$^1$, A. Lifschitz$^1$,  K. Ta Phuoc$^1$, J. Gautier$^1$, J.-P. Goddet$^1$, A. Tafzi$^1$, A. Flacco$^1$, F. Tissandier$^1$, S. Sebban$^1$,  A. Rousse$^1$, V. Malka$^1$}

%The first three authors contributed equally to this work. Correspondence and requests for materials should be addressed to C.T. (email: cedric.thaury@ensta-paristech.fr)} 

\maketitle 

\begin{affiliations}
\item Laboratoire d'Optique Appliqu\'ee, ENSTA  - CNRS UMR7639 - \'Ecole Polytechnique, 828 Boulevard des MarŽchaux , 91762 Palaiseau, France
\item Centro de Laseres Pulsados,  Parque Cient'fico, 37185 Villamayor, Salamanca, Spain
\end{affiliations}

\begin{abstract}
%Laser-plasma technology promises a drastic reduction of the size of high energy electron accelerators\cite{Nature2004Faure,Nature2004Geddes,Nature2004Mangles}. It could make free electron lasers available to a broad scientific community, and push further the limits of electron accelerators for high energy physics\cite{NatPhys2008Nakajima, NatPhys2008Malka,Schroeder2010PRSTAB}. However,  several critical issues have to be addressed before these applications become a reality. In particular, they require the electron beam to be transported, which is challenging due to the large divergence and the ultrashort duration of electron bunches from laser-plasma accelerators. Indeed, the field strength of conventional electron optics is too low to focus these electron beams while preserving their quality.  Here we show that this issue can be solved by using a laser-plasma lens,  in which  the fields can be five order of magnitude larger than in a conventional lens. We demonstrate a reduction of the divergence by nearly a factor of three, which should allow for an efficient coupling of the beam with conventional electromagnetic optics. 
Laser-plasma technology promises a drastic reduction of the size of high energy electron accelerators. It could make free electron lasers available to a broad scientific community, and push further the limits of electron accelerators for high energy physics. Furthermore the unique femtosecond nature of the source makes it a promising tool for the study of ultra-fast phenomena. However, applications are hindered by the lack of suitable lens to transport this kind of high-current electron beams, mainly due to  their divergence. 
Here we show that this issue can be solved by using a laser-plasma lens, in which the field gradients are five order of magnitude larger than in conventional optics. We demonstrate a reduction of the divergence by nearly a factor of three, which should allow for an efficient coupling of the beam with a conventional beam transport line. 
\end{abstract}

Electron beams from laser-plasma accelerators\cite{Nature2004Faure,Nature2004Geddes,Nature2004Mangles} have typical normalized transverse emittances of about or below 1~mm.mrad\cite{PRL2010Brunetti,2012PRL_Plateau,weingartner2012,moment_ang}, comparable or even smaller than those of linear accelerators delivering similar energies\cite{SPARC_2003,FLASH_emittance}.  Yet, this small emittance is mostly due  to a sub-micrometer source size\cite{corde_prl_2011a},  while the beams typically have rather large divergence of a few milliradians. Their energy spread, of a couple of percents\cite{PRL2009Rechatin1}, is also at least one order of magnitude larger than in linear accelerators. This raises several issues for the beam transport and hence for key applications of laser-plasma accelerators such as free electron lasers and high energy colliders\cite{NatPhys2008Nakajima, NatPhys2008Malka,Schroeder2010PRSTAB,PhysRevSTAB.14.091301}. In particular, the transverse  emittance tends to increase during a free drift,  because electrons with different energies rotate with different velocities in the transverse phase space\cite{Floettmann2003PRSTAB}. The emittance increase is tolerable\cite{2013PRSTAB_migliorati} if the drift length $L\lesssim\epsilon_\mathrm{N}/(\gamma_0\sigma_\mathrm{E}\sigma_\mathrm{\theta}^2)$. Here  $\epsilon_\mathrm{N}$ is the initial normalized emittance, $\gamma_0$ the mean Lorentz factor, $\sigma_\mathrm{E}$  the root-mean-square (RMS) relative energy spread and $\sigma_\mathrm{\theta}$ the RMS divergence.   Regarding state of the art laser-plasma accelerators\cite{PRL2009Rechatin1,moment_ang,buckPRL2013,2012PRL_Plateau}, the above condition indicates that the drift length  should be smaller than 1 to 5 cm, depending on the exact conditions. In other words, electrons must be focused within a few centimeters from the accelerator exit in order to be transported efficiently.

Focusing the beam within such a short distance requires very high transverse field gradients. Quadrupole magnets which are generally used to transport electron beams, operate at gradients of $\sim50$ T~m$^{-1}$, which is two orders of magnitude less than required. Though using non-adjustable permanent magnet configurations with millimeter size aperture, field gradients of up to 500 T~m$^{-1}$ have been reported\cite{Becker2009PhysRevSTAB}, further enhancement is limited due to manufacturing issues, demagnetization effects, etc. An electron beam transport line  based on quadrupole-technology will therefore degrade the quality of a laser-plasma electron beam, rendering it useless for most applications.
As they can sustain much higher gradients, plasmas could help to drastically miniaturize focusing optics, similar to the miniaturization achieved by laser plasma accelerators, and hence to avoid  any emittance growth. Incidentally, the idea to use plasma to focus an electron beam\cite{Chen1987} is almost as old as the idea to use plasma to accelerate electrons\cite{PRL1979Tajima}. It was proposed to focus an electron beam using the radial fields created in the wake of the electron beam itself, when it propagates in a plasma. The so-called plasma lens was demonstrated in the context of conventional accelerators\cite{Nakanishi1991,PhysRevLett.72.2403,Pop1995Hairapetian,PhysRevLett.87.244801}, but it has not been considered for focusing electron beams from laser plasma accelerators, due to the ultra-short length of these beams. Indeed, there is always a finite length at the bunch head over which the focusing is very non-uniform\cite{ Barov1998, Rosenzweig1991}; for ultrashort bunches from laser-plasma accelerators  this length is comparable the bunch length\cite{Lehe2014}. The laser-plasma lens was recently proposed, and validated by 3-dimensional particle-in-cell simulations,  to solve this issue\cite{Lehe2014PRSTAB2,phdlehe}. 
In the following, we present an experimental demonstration of this concept. 

%\section*{Results}
%\subsection{Principle of the laser-plasma lens}
In a laser plasma accelerator, the wakefields in which electrons are accelerated, present both longitudinal components, which are responsible for the energy gain\cite{RMP2009Esarey}, and transverse components, which make electrons oscillate and lead to betatron radiation\cite{RMP_corde}. The idea of the laser-plasma lens is to use these transverse fields to focus the electron beam. Its principle is illustrated in Fig.~1a. A laser pulse drives a wakefield in a first gas jet, diffracts in free space, and drives again a wakefield in a second jet. As a result, an electron beam is generated and accelerated in the first jet; it then drifts in the free space, where the interaction with the plasma is negligible,  and is focused by the transverse components of the wakefield in the second jet.
In general, transverse oscillations in a laser-wakefield cannot be used to focus an electron beam, because the beam electrons oscillate out of phase (there is no correlation between the electron position and its propagation angle). In a laser-plasma lens, the required synchronization is  operated by the free drift. During the drift,  the trajectory  of individual electrons is $r(z)=r_0+\theta_0 z$, with $r_0$ and $\theta_0$ being the transverse position and propagation angle at the accelerator end,   and $z$ the longitudinal  distance from the accelerator end. Hence, for $z\gg r_0/\theta_0$, $r$ is almost proportional to $\theta_0$: the position is strongly correlated to the propagation angle. Because of this correlation, electrons oscillate almost in phase in the second  laser-wakefield (the laser-plasma-lens), except for a small detuning arising from the beam energy spread (which  impacts the oscillation frequency) and  from the dependance of $r$ on $r_0$.  Thus, the plasma can act as a lens whose strength depends on its  length and density. More specifically, the electron beam will be collimated if the focusing fields vanish when the transverse momentum is minimum for most electrons.

\begin{figure*}[t!]
\begin{center}
\includegraphics[width=12cm]{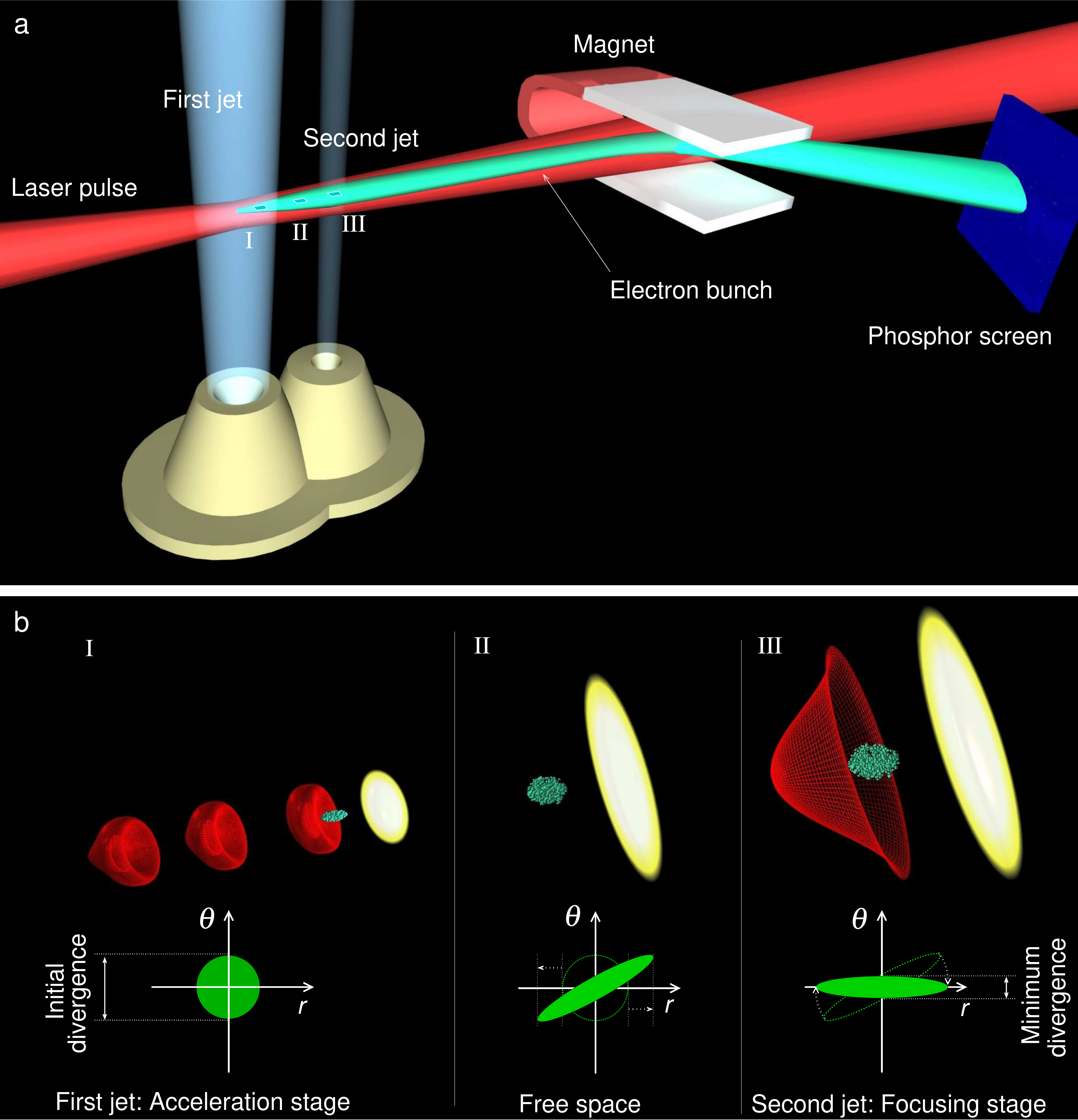}
\caption{Principle scheme of the laser-plasma lens. (a) An electron beam is accelerated in the first gas jet (accelerator), then it enters free space where it diverges, and is eventually focused in the second gas jet (lens). The same laser triggers a wakefield in both gas jets. Electron spectra are measured using an electron spectrometer consisting of a dipole magnet and a phosphor screen, imaged by a CCD camera.  (b) Phase spaces at the end of the acceleration (I), drift (II) and focusing (III) stages.}
\label{fig1}
\end{center}
\end{figure*}

The principle of the laser-plasma lens is further illustrated by the phase-space traces in Fig.~1b. The free drift creates a correlation between $r$ and $\theta$, then the phase-space trace rotates in the second gas jet, leading to a decrease of the beam divergence. The trace area gives the transverse emittance, which is conserved during the focusing process. This emittance conservation can be used to estimate the minimum divergence  $\sigma_{\mathrm{\theta,min}}\approx\epsilon_\mathrm{N}/(\gamma\sigma_\mathrm{\theta} L)$, where $\epsilon_\mathrm{N}$ and $\sigma_\mathrm{\theta}$ are the normalized emittance and the RMS divergence at the accelerator end, and $L$ is the distance between the two gas jets. We see that the drift length $L$ is a key parameter to enhance the collimation. Note that this is a rough estimate, which assumes that the transverse fields in the lens are perfectly linear and neglects both the increase of the beam size during the phase space rotation and chromatic effects, i.e. influence of the beam energy spread; a more accurate analysis shows similar trends\cite{Lehe2014PRSTAB2}.  Note also that a noticeable electron density is tolerable in the drift space as long as the transverse forces do no prevent the electron beam to diverge.  While $\sigma_\mathrm{\theta,min}$ decreases as $L$ increases, longer drift lengths do not necessarily imply smaller divergences. As the laser diffracts while propagating in free space,  the laser intensity and thus the wakefield amplitude in the second jet decrease with increasing $L$, resulting in only a \emph{partial} collimation of the electrons for long $L$. This puts an upper limit on the drift length, meaning that there is always a trade-off between low  $\sigma_\mathrm{\theta,min} $ and efficient wakefield  generation (necessary to get an actual divergence reduction).

%\subsection{Influence of the drift length}

The laser-plasma lens, in the collimation configuration, was demonstrated at LOA using the Salle Jaune laser system (see Methods, for information on the laser system, the gas targetry and  divergence measurements). At  the end of the accelerator stage, electrons had a mean energy of $\approx 241\pm 12 $ MeV (RMS error) and a beam divergence of $\approx 4.1\pm 0.6$ mrad.  The electron beam had a few pC charge and was stable shot-to-shot, suggesting that electrons are injected into the accelerator by longitudinal self-injection\cite{corde2013natcom}. The laser plasma lens does not modify significantly the charge, the mean energy nor the spectrum shape (see Fig.~2a). In contrast, Fig.~2b shows that  the beam divergence can strongly decrease in the lens, down to $1.6$ mrad. The final divergence depends on the distance $L$ from the accelerator to the lens. It decreases as $L$ increases for $L\lesssim 1.8$ mm  and increases with $L$ for $L\gtrsim 1.8$ mm. The evolution of the divergence is governed by two effects, the decrease of the minimum achievable divergence  $\sigma_{\theta,min}$,  and  the decrease of the laser intensity, which reduces the strength of the focusing fields for long $L$.  Indeed, the Rayleigh length at the accelerator exit is below 400 $\mu$m (see Methods). After 1.8 mm of propagation the laser beam diameter and intensity are thus $\Phi>54$ $\mu$m FWHM and $I < 8.3 \times10^{17}$~Wcm$^{-2}$.  Therefore, the wakefield in the lens is in the linear regime and the strength of the focusing fields strongly depends on the laser beam size and intensity.  More precisely,  the gradient of the focusing fields in the lens evolves as\cite{Lehe2014PRSTAB2}  $z^{-4}$. As a result, these fields can be too strong for short $L$, leading to excessive focusing, and too low for long $L$,  leading to insufficient focusing. In Fig.~2, the transverse force seems to vanish for $L\gtrsim4.5$ mm, while the lowest divergence is obtained for $L\approx 1.8$ mm.  Note that for the shortest lengths the wakefield may be in the non-linear regime.

\begin{figure}[t!]
\begin{center}
\includegraphics[width=85mm]{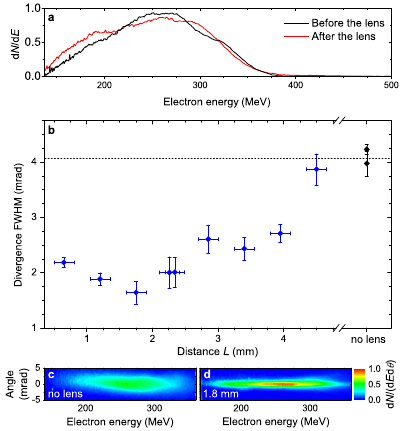}
\caption{Influence on the electron beam divergence of the distance between the accelerator and the lens $L$. (a) Electron spectrum before and after the lens, normalized by $1.3\times 10^5$. The distance between the two stages is $L=1.8$ mm. (b) Full-width-at-half-maximum (FWHM) beam divergence, at $270$ MeV, as a function of $L$ (blue diamonds). The divergence measured without the laser-plasma lens is indicated by the black diamonds on the right.  The peak electron density in the focusing stage is about $3.9\times10^{18}$ cm$^{-3}$. Data points were averaged over 10 shots; the  vertical error bar corresponds to the standard error of the mean, and the horizontal bar the precision on the measurement of $L$ . Two series of ten shots were fired for $L=2.3$ mm, and the case without lens. (c) Typical angularly resolved spectrum without laser-plasma lens. (d) Typical spectrum obtained for $L=1.8$ mm. The color scale indicates the number of electrons per MeV and mrad, divided by 17000.  }
\label{fig2}
\end{center}
\end{figure}

%\subsection{Influence of the plasma-lens density}

In addition to the drift length, the electron density $n_2$ and the length of the second jet can also impact the lens properties. The influence of the  electron density is illustrated in Fig.~3. 
According to the model from Ref.~[29] the focusing force varies as $\left[ k_\mathrm{p}\sigma_\mathrm{z} \exp(-k_\mathrm{p}^2\sigma_\mathrm{z}^2/2)\right]\sin(k_\mathrm{p}d)$, where $k_\mathrm{p}\propto n_2^{1/2}$ is the plasma wave vector, $\sigma_\mathrm{z}$ the RMS length of the laser pulse and $d$ the distance between the laser and the considered electron. The term in brackets corresponds to the amplitude of the transverse wakefield; it is maximum when the laser pulse is resonant with the plasma wave, \emph{i.e.}  for $k_\mathrm{p}\sigma_\mathrm{z}=1$. The strength of the lens depends also on the position of the electron beam in the wakefield. This effect is described by the sine term; electrons experience the largest focusing force when $k_\mathrm{p}d=\pi/2$, and are defocused for $k_\mathrm{p}d>\pi$. The combination of both terms leads to a complex influence of the electron density on the focusing strength. 
For low densities ($n_2\lesssim 1\times10^{18}$ cm$^{-3}$ in Fig.~3),  the focusing fields are very weak and the divergence is hardly reduced. As density increases the transverse focusing fields rise, which leads to the desired beam collimation. In Fig.~3, the lowest divergence is obtained for $n_2\approx 4.3\times10^{18}$ cm$^{-3}$. The fact that the divergence remains almost constant for higher densities suggests that the focusing force has a local maximum, in the range $n_2\lesssim6.2 \times10^{18}$ cm$^{-3}$, around $n_2\approx 4.3\times10^{18}$ cm$^{-3}$. For even higher densities (not investigated in the experiment), the divergence should increase and eventually exceed the initial divergence when the fields become defocusing (for $ k_\mathrm{p}d>\pi$). Note that, for long $L$, the mean density which is effectively experienced by the electron beam during the focusing process  may be significantly smaller than the peak density indicated in Fig.~3, because the focusing process is likely to occur only in the rising edge of the second gas jet (the focusing force decreases as $z^{-4}$).

\begin{figure}
\begin{center}
\includegraphics[width=85mm]{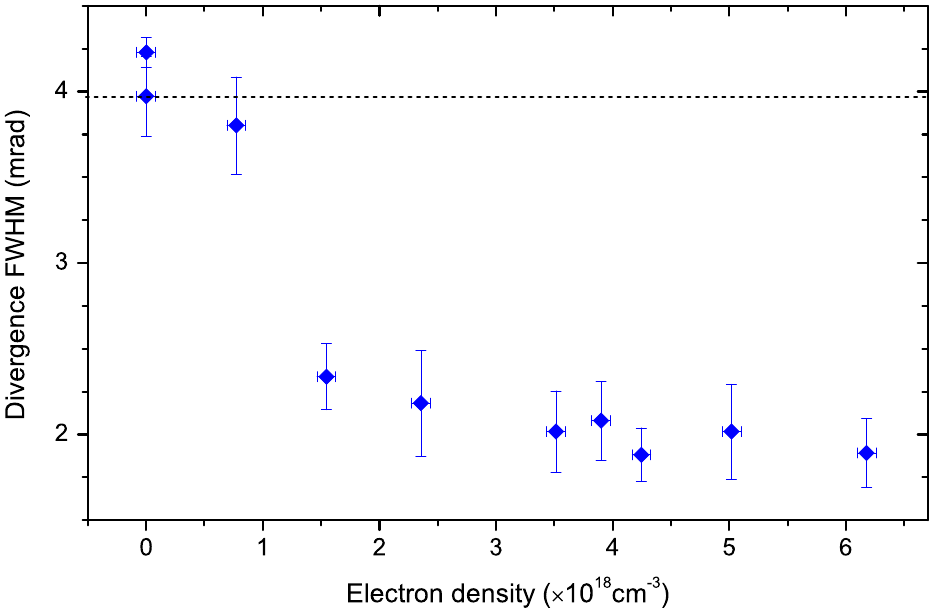}
\caption{Influence on the electron beam divergence of the peak electron density in the second gas jet (the laser-plasma lens). The divergence was measured for an electron energy of $270$ MeV. The distance  between the two stages is $L=2.3$ mm. Data points were averaged over at least 11 shots; the vertical error bar correspond to the standard error of the mean, and the horizontal bar to the precision on the backing pressure. The analysis of the interferograms leads to an additional  systematic error on the density of $\pm 5\%$.}
\label{fig3}
\end{center}
\end{figure}

%\subsection{Chromaticity of the lens}
Up to this point, the electron beam was implicitly assumed to be mono-energetic, and in Figs.~2b and~3 only the divergence at 270 MeV was considered. Yet,  a careful analysis of the angularly-resolved electron spectra in Fig.~2c-d reveals that electrons of different energies are not focused to the same extend. This is shown more clearly in Fig.~4, which displays the reduction factor of the beam divergence as a function of both the electron energy and the distance between the two stages. A couple of observations can be made from this plot. First, the reduction factor is larger towards higher energies. This is likely due to the decrease of  $\sigma_\mathrm{\theta,min}$ at higher energies, both because  $\sigma_\mathrm{\theta,min}$ varies as  $(\gamma\sigma_\mathrm{\theta} )^{-1}$, and because in this particular experiment, the divergence $\sigma_\mathrm{\theta} $ at the accelerator end increases with $\gamma$. Second, the optimal length $L$ decreases as the electron energy increases. This is because the amplitude of the transverse fields required to focus the beam increases with the electron energy. Lengthening  $L$ tends to lessen the focusing fields and hence enhances the focusing of low energy electrons (similar results can be obtained by decreasing the electron density).
 Nonetheless, Figs.~2b-c and~4 show that an electron beam with an energy spread exceeding $100$ MeV can be strongly focused, as a whole, using a laser-plasma lens.

\begin{figure}
\begin{center}
\includegraphics[width=85mm]{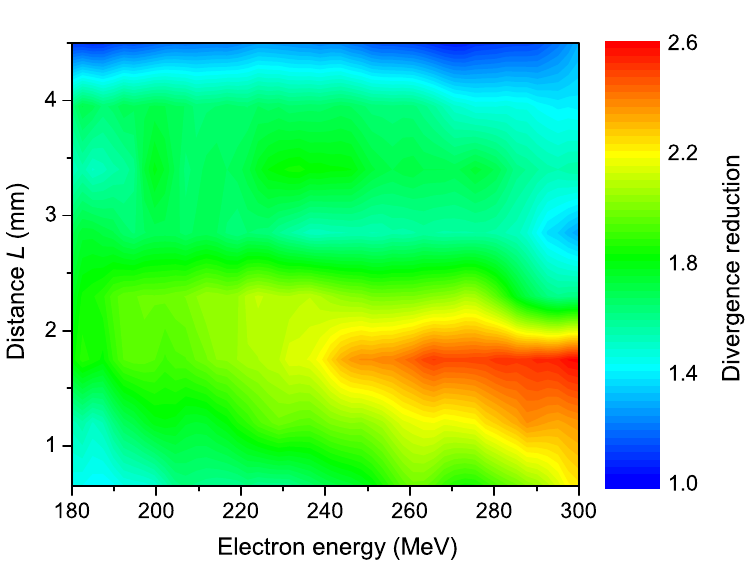}
\caption{Chromaticity of the laser-plasma lens. The color-map shows the factor of reduction of the divergence, as a function of the electron energy and of the distance $L$ between the accelerator and the lens.  The peak electron density in the focusing stage is about $3.9\times10^{18}$ cm$^{-3}$.}
\label{fig4}
\end{center}
\end{figure}

%\section*{Discussion\\}

We demonstrated the collimation of an electron beam by a laser-plasma lens. The electron beam was accelerated in a laser-plasma accelerator up to  $\approx 300$ MeV, before going through a laser-plasma lens. At the lens exit, the beam divergence was reduced by a factor of 2 for the whole beam, and a factor of 2.6 for its high energy part. This factor  was  limited by the fast decrease of the laser intensity in the lens. Stronger collimation could be obtained by using a shorter gas jet for the acceleration stage, a gas jet with sharper gradients for the lens, or a more energetic laser pulse.
Nevertheless, the demonstrated divergence reduction should already be sufficient to transport the electron beam with a quadrupole lens. For state of the art laser-plasma accelerators, the emittance growth should remain negligible over a propagation distance of about 30 cm, which should be enough to transport the beam with compact quadrupoles\cite{Becker2009PhysRevSTAB}.   Alternatively, the collimated electron beam could be send directly  into an optical\cite{andriyashPRL2012} or a plasma undulator\cite{andriyashNatCom2014}  to form a millimeter scale, possibly coherent,  synchrotron source.

  \begin{methods}
  	\subsection{Laser system}
	The experiment was conducted at Laboratoire d'Optique Appliqu\'ee (LOA) with the ``Salle Jaune" Ti:Sa laser system, which delivers 0.9 Joule  in the focal spot with a full width at half maximum (FWHM) pulse duration of 28 fs and a linear polarization. The laser pulse was focused at the entrance of the first gas jet with a 1 m focal length off-axis parabola, to a FWHM focal spot size of 12 $\mu$m. From the measured intensity distribution in the focal plane, the peak intensity was estimated to be $1.8\times10^{19}$ Wcm$^{-2}$. The Rayleigh length in vacuum is about 400 $\mu$m. It is likely significantly shorter at the accelerator exit because of self-focusing.

	\subsection{Gas targetry}
	Helium gas jets were used for both the acceleration and the focusing stages. The two gas jets were produced by supersonic nozzles. The first has an exit  diameter of 3 mm and a Mach number of 3, the second has a diameter of 0.8 mm and a Mach number of 1.6. The laser was fired at  1.6 mm from the nozzle exits.The electron density profiles of both gas jets were characterized by interferometry, and analyzed using the image processing program Neutrino. 
The density profile in the acceleration stage had a plateau of $2.4\pm 0.1$ mm surrounded by $600 \pm 100$ $\mu$m gradients. The peak electron density was about $9.2\pm 0.5\times10^{18}$ cm$^{-3}$. In the second gas jet, the density profile was triangular with $1\pm0.1$ mm gradients. The distance between the two gas jets was measured at half maximum.
		
	\subsection{Electron diagnostics}
	The electron divergence was measured, in the vertical direction, from electron spectra. The electron spectrometer consisted of a permanent bending magnet (1.1 T over 10 cm) which deflects electrons depending on their energy, and a Lanex phosphor screen which converts a fraction of the electron energy into 546 nm light imaged by a 16 bit visible CCD camera. The energy resolution varies between 1\% (for 140 MeV electrons with a beam divergence of 1.5 mrad)  and 10\% (for 300 MeV electrons with a beam divergence of 4 mrad). The angular resolution is about 0.3 mrad.
Two dimensional footprints of the electron beam showed that the divergence is very similar in the horizontal and vertical directions. These footprints cannot be used to estimate the divergence reduction, because of chromatic effects.

  \end{methods}

{\textbf{Acknowledgments}}
\noindent
This work was supported by the European Research Council through the  ERC projects  PARIS (Contract No. 226424) and X-Five (Contract No. 339128), and by the Agence Nationale pour la recherche through the 
project LUCEL-X (ANR-13-BS04-0011). A.D. acknowledges LA3NET, which is funded by the European Commission under Grant Agreement Number GA-ITN-2011-289191. C. T. thanks A. Loulergues for fruitful discussions regarding the emittance growth in beam transport lines. %and CILEX (ANR-10-EQPX-CILEX)

{\textbf{Author contributions}}
\noindent
C.T. proposed the experiment and wrote the paper with help from E.G, A.D. and R.L.. E.G., A.D., C.T., K.T.P. and J.G. performed the experiment. E.G. analyzed the results. R.L. and A.L. provided a theoretical support. A.L realized Fig. 1. J.P.G. and A.T. designed, built and operated the upgraded laser system of ÒSalle JauneÓ. A.F., F.T., S.S. and A.R.  provided support for the operation of the facility. V.M supervised the project.

\end{document}